\documentclass[12pt]{iopart}
\usepackage[utf8]{inputenc}
\begin{document}

\title{General Null Lagrangians and Their Novel Role in Classical Dynamics}

\author{R. Das$^1$ and Z.E. Musielak$^2$} 

\address{$^1$Department of Physics and Earth Space Science, The University 
of Indianapolis, Indianapolis, IN 46227, USA}
\address{$^2$Department of Physics, The University of Texas at Arlington, 
Arlington, TX 76019, USA}

\ead{dasr@uindy.edu; zmusielak@uta.edu}

\begin{abstract}
A method for constructing general null Lagrangians and their higher 
harmonics is presented for dynamical systems with one degree of 
freedom.  It is shown that these Lagrangians can be used to obtain 
non-standard Lagrangians, which give equations of motion for the 
law of inertia and some dissipative dynamical systems.  The necessary 
condition for deriving equations of motion by using null Lagrangians 
is presented, and it is demonstrated that this condition plays the same 
role for null Lagrangians as the Euler-Lagrange equation plays for 
standard and non-standard Lagrangians.  The obtained results and 
their applications establish a novel role of null Lagrangians in classical 
dynamics.  
\end{abstract}

\maketitle

\section{Introduction} 

There are three families of Lagrangians that can be classified as standard, 
non-standard and null Lagrangians.  The main characteristic of standard 
Lagrangians is the presence of the kinetic and potential energy-like terms, 
and these Lagrangians are commonly used in Classical Mechanics (CM) to 
derive equations of motion for different dynamical systems (e.g., [1-5]).  
On the other hand, in non-standard Lagrangians, neither the kinetic nor 
the potential energy-like terms can be identified.  Different methods have 
been developed to derive the standard and  non-standard Lagrangians for 
given ordinary differential equations (ODEs) [6-12]. 

The third family of null Lagrangians (NLs) has the following two main equivalent
characteristics: (i) they must satisfy identically the Euler-Lagrange (E-L) 
equation: the null condition, and (ii) they must be expressed as the total derivative of any 
scalar function known as gauge function: the gauge condition.  The properties and 
applications of these null Lagrangians have been extensively explored 
in different fields of mathematics (e.g., 13-20]) as well as in some 
physical applications (e.g., [21,22]) that include restoring Galilean 
invariance of Lagrangians in Newtonian dynamics [23,24], and 
introducing forces to CM [25-27].

Several methods of constructing the NLs have been proposed and 
most of them rely on specifying a gauge function and using it to 
obtain the resulting null Lagrangian [13,14,26,27].  However, in 
this paper, we develop a new method that is based on a generating 
function that differs from the gauge function.  It is shown that using 
this generating function a general null Lagrangian can be obtained.
Depending on the form of the generating function, either standard 
or  non-standard null Lagrangians can be constructed [28].  Moreover,
the method also allows us to derive higher harmonics of these null 
Lagrangians, and the harmonics are new mathematical objects in 
calculus of variations; therefore, their role and possible applications 
are discussed.

Since the NLs satisfy identically the E-L equation, they can be added 
to any standard or  non-standard Lagrangian, and this addition does 
not affect the resulting equation of motion [13-18].  Despite this 
well-known property of the NLs, we demonstrate that they may be 
used to construct non-standard Lagrangians, which then give equations 
of motion when susbtituted into the E-L equations.  Using this approach,
the NLs for some equations of motion can be found and compared to 
other Lagrangians for these equations, which may give new insights 
into symmetry of the Lagrangians and the resulting equations of motion 
[29-33].  The fact that equations of motion can be derived by using the 
NLs demonstrates a novel role of these Lagrangians in classical dynamics.

Applications of the obtained results show that the law of inertia can be 
derived from its null Lagrangian, and that other constructed NLs give
equations of motion for some dissipative systems with linear and 
quadratic dissipative terms.  There have been many attempts to 
establish Lagrangian formalism for dissipative systems [7-9,27,34-47].
However, none of the previous work on this problem has involved 
null Lagrangians and shown that they can be used to obtain equations 
of motion.  The results of this paper demonstrate the validity of this 
approach and presents the necessary condition that null Lagrangians 
must obey in order to give equations of motion.  The condition plays 
the same role for null Lagrangians as the Euler-Lagrange equation plays
for standard and non-standard Lagrangians.
 
The paper is organized as follows.  In Section 2, Lagrangian formalism 
and its Lagrangians are briefly described; construction of null Lagrangians
and  non-standard Lagrangians resulting form them is presented in Sections
3 and 4, respectively; applications of the results to dynamical systems are 
considered and discussed in Section 5; and conclusions are given in Section 6.

\section{Lagrangian formalism and its Lagrangians}
\label{formalism}

\subsection{Standard and non-standard Lagrangians}

Dynamical systems with holonomic constraints can be analyzed using the Lagrangian
formalism. The foundation of this formalism is the smooth configuration manifold
$\mathcal{Q}$ constructed from the generalized coordinates of the system of interest 
with holonomic constraints and its tangent bundle $T\mathcal{Q}$.  A Lagrangian 
$\mathcal{L}$ for this system is commonly defined as a $C^{\infty}$ map from 
$T\mathcal{Q}$ to the real line $\mathbf{R}$, i.e., $\mathcal{L}:T\mathcal{Q}
\rightarrow \mathbf{R}$ [3]. 

In the case of one-dimensional dynamical systems, a Lagrangian can be symbolized 
as $\mathcal{L} = \mathcal{L}(x,\dot{x})$ in an open neighborhood $\mathcal{U}
\subset T\mathcal{Q}$ containing a point $p\in T\mathcal{Q}$ in a local coordinate 
system defined by the coordinate function $x:\mathcal{Q}\rightarrow \mathbf{R}$ 
and its associated tangent vector $\dot{x}$ at the point $p$  using a diffeomorphism 
$\phi :p\in \mathcal{U}\subset{T{\mathcal{Q}}} \rightarrow \mathcal{Q} \times 
\mathbf{R}$ known as a local trivialization.  Similarly, the explicit dependence of 
a Lagrangian on time can be defined as $\mathcal{L}:T\mathcal{Q}\times \mathbf
{R}\rightarrow \mathbf{R}$ and thus expressing $\mathcal{L} = \mathcal{L}(x,
\dot{x},t)$. After a suitable Lagrangian for an interested system is constructed or 
discovered, the time evolution of the system is analyzed by applying the Euler-Lagrange 
operator on the chosen Lagrangian, i.e., $\hat {EL} [ \mathcal{L} (\dot x, x, t) ] = 0$, 
with $\hat {EL}$ being the Euler-Lagrange (E-L) operator [2-4].

The Euler-Lagrange (EL) operator in each degree of freedom for a system is a linear 
operator acting on the vector space of $C^{\infty}$ functions defined on $T\mathcal{Q}$ 
and thus  $\mathcal{L}\in \mathcal{F}(T\mathcal{Q})$ with $\mathcal{F}$ denoting 
the space of $C^{\infty}$ functions.  Most commonly, a standard strategy for the 
construction of the most common Lagrangians (thus known as standard Lagrangians) 
are carried out from the difference of the kinetic and potential energy of the respective 
system under consideration, i.e., $\mathcal{L}(x, \dot{x}) = \frac{1}{2} \dot{x}^{2}-
V(x)$, with the potential energy $V(x)$ for a system with only one degree of freedom [1-3]. 

On the other hand, non-standard Lagrangians are dependent on functions of time in 
addition to the usual dynamical variables such as $x$ and $\dot{x}$ for systems with 
only one degree of freedom [7-11].  In general, the construction of these non-standard 
Lagrangians for the set of desired equations of motion requires solving nonlinear 
(Riccati-type) equations [9,12].  However, the problem can be simplified by replacing 
functions to be determined by constants, for instance, we may consider the following 
form of the non-standard Lagrangian 
\begin{equation}
\mathcal{L}_{ns} (\dot x , x , t)  = {1 \over {a_1 \dot{x}  + a_2 t 
+ a_3}}\ ,
\label{test-non-standard}
\end{equation}
with constants $a_1$, $a_2$, and $a_3$ describes a system undergoing a constant 
acceleration. It should be emphasized that the above method for constructing non-standard 
Lagrangians $C^{\infty}$ functions demands an extra constraint that the denominators 
do not vanish, i.e., ${a_1 \dot{x}  + a_2 t + a_3}\neq 0$ in the domain $\mathcal{U}
\subset{T\mathcal{Q}}$. Therefore, non-standard Lagrangians in general represent 
dynamical systems with more constraints than the usual accompanying holonomic constraints. 
The construction of these Lagrangians can be carried out with the assumption that the 
corresponding Lagrangians are well-behaved on the domain $\mathcal{U}\subset{T\mathcal{Q}}$.  

\subsection{Null Lagrangians}

Since the null space of the Euler-Lagrange (EL) operator is itself a vector subspace, the 
members of this null space, rightly known as the null Lagrangians, are nullified by the 
E-L operator. More explicitly, an action functional defined in terms of null Lagrangians as 
\begin{equation}
A [x ; t_e, t_o] = \int^{t_e}_{t_o} L_{\rm null} (t, x, \dot{x})dt\ ,
\label{null-L}
\end{equation}
satisfies the property $A [x ; t_e, t_o] =A [x +\eta; t_e, t_o]$ for continuous functions 
$x(t)$ and $\eta (t)$ of time on the same domain. This implies through approximations
[13,14] that $A [x ; t_e, t_o] =A [\eta; t_e, t_o]$ for all permissible continuous functions 
$x(t)$ and $\eta (t)$ on the domain such that $x(t_e)=\eta (t_e)$ and $x(t_0)=\eta (t_o)$. 
In other words, the above action integral produces the same real number for all admissible 
trajectories $x(t)$'s between $t_e$ and $t_o$ in the same domain, which in turn implies that 
the variation of the action integral is rendered identically to zero. This also implies that no 
preferred trajectory that extremizes the action integral exists for the relevant null Lagrangians. 
Therefore, the E-L operator renders all the null Lagrangians defined on the same domain 
identically to zero, i.e., no equation of motion can be obtained directly from the application 
of the E-L operator on null Lagrangians.

However, the study of null space of the E-L operator has been of particular interest due 
to its applicability to the exploration of symmetries in field theories. An effort to construct 
these null Lagrangians has been an active practice for a while [13,14]. The most prevalent 
constructions centered around the observation that the E-L operator nullifies the total time 
derivative of an arbitrary $C^1$ (continuous first derivatives) function $\Phi (x,t)$, known 
as the gauge function, defined on the domain of interest $\mathcal{U}\subset{T\mathcal{Q}}$. 
More explicitly, a Lagrangian is null if, and only if, it is the total time derivative of a scalar 
function $\Phi(x,t)$ [13] that can be expressed as
\begin{equation}
L_{null}(\dot{x},x,t)=\frac{d\Phi (x,t)}{dt}=\frac{\partial \Phi}{\partial x} \dot{x} + 
\frac{\partial \Phi}{\partial t}.
\label{null-gauge}
\end{equation}
Again, it should be emphasized that the above form of a null Lagrangian is rendered identically 
to zero by  the E-L operator, i.e., $\hat{EL}[L_{null}:=\frac{d\Phi (x,t)}{dt}]=0$. Then a 
natural strategy to construct a standard null Lagrangian is to cast it in the following form:
\begin{equation}
L_{null}(\dot{x},x,t)= B(x,t) \dot{x}+ C(x,t)  
\label{null-Lag}
\end{equation}
such that $B(x,t):=\frac{\partial \Phi}{\partial x}  $ and $C(x,t):=\frac{\partial \Phi}{\partial t} $. 
A similar construction has been presented in [14].  A simple example of a standard null 
Lagrangian can be obtained from a given gauge function $\Phi(x,t):=f_{1}(t)x^{2}+f_{2}(t)$ as 
$L_{null}=2f_{1}(t)x\dot{x}+\dot{f}_{1}(t)x^{2}+\dot{f}_{2}(t)$ with the obvious identifications 
$B(x,t)=2f_{1}(t)x$ and $C(x,t)= \dot{f}_{1}(t)x^{2}+\dot{f}_{2}(t)$. It is also possible to 
construct non-standard null Lagrangians following the same method as above. For example, 
the following non-standard null Lagrangian
\begin{equation}
L_{ns,test1} (\dot x , x , t)  = {{a_1 \dot x } \over {a_2 x 
+ a_4}}\ .
\label{S3eq2}
\end{equation}
can be obtained from $\Phi ( x, t ) = {{a_1} \over {a_2}} \ln \vert a_2 x  + a_4 \vert$ [28]. 
Therefore, it is always possible to construct a standard as well as a non-standard null Lagrangian 
in the above form in Eq.(\ref{null-Lag}) if a corresponding gauge function $\Phi(x,t)$ is known. 

On the contrary, it is also possible to construct some standard null Lagrangians in the above form 
in Eq. (\ref{null-Lag}) without any reference to the corresponding gauge functions.  In the following 
section, an earnest humble effort is made to construct a set of both standard and non-standard null 
Lagrangians and their higher harmonics in the similar form as in Eq. (\ref{null-Lag}) without explicitly 
resorting to any gauge function.  

\section{Method to construct null Lagrangians}
\label{construction}

\subsection{Standard null Lagrangians}

The construction of null Lagrangians is motivated by the recognition of the null-condition 
$\hat {EL} [ \mathcal{L}_{null} (\dot x, x, t) ] = 0$, i.e., the action of the E-L operator 
on these smooth functions nullifies them identically to zero.  In other words, these null 
Lagrangians must satisfy the following relation:
\begin{equation}
    \frac{dp_{null}}{dt}= \frac{\partial L_{null}}{\partial x}\ ,
\label{null-condition}
\end{equation}
with the generalized momentum $p_{null} = \partial{L_{null}}/\partial{\dot{x}}$. 

{\bf Proposition 1:} Let $B(x,t)$, $C(x,t)$ and $f(t)$ be locally $C^1$ functions 
on $\mathcal{U}\subset{T\mathcal{Q}}$ such that they satisfy the following condition:
\begin{equation}
\frac{d B(x,t)}{d t} = \left[\frac{\partial B(x,t)}{\partial x}\right]\dot{x}+ \frac{\partial 
[xC(x,t)]}{\partial x}.
\label{BC-condition}
\end{equation}
 Then $B(x,t)$ generates a family of null Lagrangians $L_{null}(\dot{x},x,t)$ of the form 
\begin{equation}
    L_{null}(\dot{x},x,t) = B(x,t)\dot{x}+C(x,t)x+f(t).
\label{null-lagrangians}
\end{equation}
{\bf Proof:} The proof follows from the null-condition $\hat {EL} [ {L}_{null} (\dot x, x, t) ] 
= 0$ and Eq. (\ref{null-condition}).

It is interesting to note that the only restriction on $B(x,t)$ and $C(x,t)$ is that they be 
continuously differentiable functions on $\mathcal{U}\subset{T\mathcal{Q}}$ and they 
satisfy the null condition, i.e., Eq. (\ref{null-condition}). A similar construction is presented 
in [14] albeit with a different constraint.  

However, the application of the null condition, Eq. (\ref{BC-condition}), on a Lagrangian of the form in Eq. (\ref{null-lagrangians}) implies that there exists a gauge function $\Phi(x,t)$ such that the gauge condition, $L_{\rm null}(\dot{x}, x,t)=d\Phi (x,t)/dt$, is automatically satisfied. Therefore, the null condition on the function $B(x,t)$ is equivalent to the gauge condition on the gauge function $\Phi(x,t)$ related to the same null Lagrangian. In this paper, no direct reference to a gauge function $\Phi(x,t)$, i.e., no a priori assumption of the gauge condition on the gauge function is made to construct the corresponding null Lagrangian. Instead, a general function $B(x,t)$, called a generating function, is chosen along with its null condition to construct the appropriate null Lagrangian.
 Now, some choices of $B(x,t)$ generating different null 
Lagrangians would clarify the applications of the above proposition. 
\begin{enumerate}

\item

For the simplest choice of a constant function $B(x,t)$, Eq. (\ref{BC-condition}) forces both 
$xC(x,t)$ and $L_{null} (\dot{x},x,t)$ to be functions of $t$ only, implying that $\hat {EL} 
[ {L}_{null} (\dot x, x, t) ] = 0$.

\item

If $B(x,t)$ is a linear function of $x$, then the null-Lagrangian presented in [27] is 
reproduced.  Let $B(x,t)=f_{1}(t)x+f_{2}(t)t+f_{3}(t)$ such that $f_1 (t)$, 
$f_2 (t)$, $f_3 (t)$ and $f_4 (t)$ are assumed to be arbitrary but at least twice 
differentiable functions of the independent variable, then the following null 
Lagrangian [26] is obtained
\[
{L}_{null1} (\dot x , x , t)  = \left[f_{1}(t)x+f_{2}(t)t+f_{3}(t)\right] \dot{x}
\]
\begin{equation}
\hskip0.25in + \left[\frac{1}{2}\dot{f}_{1}(t)x+\dot{f}_2 (t) t  + f_{2}(t)   + 
\dot f_3 (t) \right ]x +  f_4 (t)\ .
\label{standard-null}
\end{equation}

\item

Now, let $B(x,t)=f_{1}(t)x^2+f_{2}(t)t+f_{3}(t)$. Then the corresponding null 
Lagrangian turns out to be
\[
{L}_{null2} (\dot x , x , t)  = \left[f_{1}(t)x^2+f_{2}(t)t+f_{3}(t)\right] \dot{x}
\]
\begin{equation}
\hskip0.25in + \left[\frac{1}{3}\dot{f}_{1}(t)x^{2}+\dot{f}_2 (t) t  + f_{2}(t)   
+ \dot f_3 (t) \right ]x +  f_4 (t)\ .
\label{standard-null-1}
\end{equation}

\item

It follows from the above special choices of $B(x,t)$ that it can be any polynomial of $x$. 
However, in general, $B(x,t)$ can be any continuously differentiable function of any order 
such as a $C^k$ function to obtain a null Lagrangian.  To show that, let $B(x,t)=f_{1}(t)
\sin(x)+f_{2}(t)e^{x}t+f_{3}(t)$. Then, the corresponding null Lagrangian turns out to be
\[
{L}_{null3} (\dot x , x , t)  = \left[f_{1}(t)sin(x)+f_{2}(t)e^{x}t+f_{3}(t)\right] \dot{x}
\]
\begin{equation}
\hskip0.25in -\dot{f}_{1}(t)cos(x)+\dot{f}_2 (t)e^{x} t  + f_{2}(t)e^{x}   + \dot f_3 (t)x + 
f_4 (t)\ .
\label{standard-null-2}
\end{equation}

\end{enumerate}

Furthermore, since the addition of a total time derivative of a function whose variation 
vanishes at the both ends of the trajectory does not affect the dynamics of the system, 
the total time derivative of $B(x,t)$ can be added to the null-Lagrangian in Eq. 
(\ref{null-lagrangians}) to generate its higher harmonics, which is presented below 
in Proposition 2.

{\bf Proposition 2:} Let $B(x,t)$, and $C(x,t)$ be locally $C^{\infty}$ functions on 
$\mathcal{U}\subset{T\mathcal{Q}}$ such that they are differentiable up to order $n$
 to satisfy the following condition:
\begin{equation}
\frac{d B_{n}(x,t)}{d t} = \left[\frac{\partial B_{n}(x,t)}{\partial x}\right]\dot{x}+ \frac{\partial 
[xC(x,t)]_{n}}{\partial x},
\label{BC-condition-harmonics}
\end{equation}
with

\begin{equation}
B_{n}(x,t):=\Sigma^{n}_{i=0}\mathcal{C}^{n}_{n-i}B^{(i)}(x,t)
\label{B_n}
\end{equation}
and
\begin{equation}
[xC(x,t)]_{n}:=\Sigma^{n}_{i=0}\mathcal{C}^{n}_{n-i}\left[xC(x,t)
\right]^{(i)}.
\label{C_n}
\end{equation}
Here, $(i)$ represents the $i$-th spatial derivative and $\mathcal{C}^{n}_{n-i}
=n!/[(n-i)!i!]$. Then $B(x,t)$ generates a family of higher harmonics of null 
Lagrangians $L_{null}(\dot{x},x,t)$ of the form 
\begin{equation}
 L^{(n)}_{null}(\dot{x},x,t)  =
  B_{n}(x,t)\dot{x}+
\left[xC(x,t)\right]_{n}+f(t).
\label{null-lagrangians-harmonics}
\end{equation}

{\bf Proof:} Starting from the base case, i.e., the first harmonic, $L^{(1)}_{null}=
L_{null}+\frac{dB(x,t)}{dt}$, it follows by induction that $L^{(n)}_{null}=
L^{(n-1)}_{null}+\frac{dB_{n-1}(x,t)}{dt}$. Then the application of the null 
condition for the higher harmonics, i.e., Eq. (\ref{BC-condition-harmonics}), 
proves equation Eq. (\ref{null-lagrangians-harmonics}).

Again, if $B(x,t)$ is a linear function of $x$, then the null-Lagrangian presented in 
[26,27] is reproduced.  Let $B(x,t)=f_{1}(t)x+f_{2}(t)t+f_{3}(t)$ and the following 
harmonics of the null-Lagrangian in Eq. (\ref{standard-null}) are obtained
\[
{L}^{(1)}_{null1} (\dot x , x , t)  = \left[f_{1}(t)(x+1)+f_{2}(t)t+f_{3}(t)\right] \dot{x}
\]
\begin{equation}
\hskip0.25in + \dot{f}_{1}(t)(\frac{x}{2}+1)x+\left[\dot{f}_2 (t) t  + f_{2}(t)   + 
\dot f_3 (t) \right ](x+1) +  f_4 (t)\ ,
\label{standard-null-1'}
\end{equation}
and the highest harmonics
\[
{L}^{(2)}_{null1} (\dot x , x , t)  = \left[f_{1}(t)(x+2)+f_{2}(t)t+f_{3}(t)\right] \dot{x}
\]
\begin{equation}
\hskip0.25in + \dot{f}_{1}(t)(\frac{x^2}{2}+2x+1)+\left[\dot{f}_2 (t) t  + f_{2}(t)   
+ \dot f_3 (t) \right ](x+2) +  f_4 (t) .
\label{standard-null-11'}
\end{equation}

Let $B(x,t)=f_{1}(t)sin(x)+f_{2}(t)e^{x}t+f_{3}(t)$. Then, the corresponding null 
Lagrangian turns out to be
\[
{L}^{(1)}_{null3} (\dot x , x , t)  = \left[f_{1}(t)[sin(x)+cos(x)]+2f_{2}(t)e^{x}t+
f_{3}(t)\right] \dot{x}
\]
\begin{equation}
\hskip0.25in \dot{f}_{1}(t)[sin(x)-cos(x)]+2\dot{f}_2 (t)e^{x} t  + 2f_{2}(t)e^{x}  
+ \dot f_3 (t)(x+1) +  f_4 (t),\ 
\label{standard-null-3'}
\end{equation}
and the higher harmonics can be computed according to Eq. (\ref{null-lagrangians-harmonics}).
It turns out that the above strategy can also be employed to generate a family of non-standard 
null Lagrangians as presented below.

\subsection{Non-standard null Lagrangians}

The key idea to generate a non-standard null-Lagrangian is to make a judicious choice of $B(x,t)$ 
as a fraction defined as $B(x,t)= f(t)/h(x,t)$ with $h(t)$ being a non-vanishing differentiable and 
integrable function. Once a choice on $B(x,t)$ is made, the corresponding non-standard null 
Lagrangian can be obtained by following the procedure described above in Proposition 1.   
Let a generating function $B(x,t)$ be expressed as
\begin{equation}
    B(x,t) = \frac{f_{1}(t)}{f_{2}(t)x+f_{3}(t)t+f_{4}(t)},
    \label{generating-function}
\end{equation}
where $f(t)$'s are differential functions of time and $f_{2}(t)x+f_{3}(t)t+f_{4}(t) \ne 0$. 
Then, the corresponding non-standard null Lagrangian can be obtained by using Eqs. 
(\ref{BC-condition}) and  (\ref{null-lagrangians}) as
\[
L_{nsn}(\dot{x}, x,t) = \frac{f_{1}(t)\dot{x}}{f_{2}(t)x+f_{3}(t)t+f_{4}(t)}
\]
\[
\hskip0.25in +\frac{h_{2}(t)}{f_{2}^{2}}\left[ln \vert f_{2}x+f_{3}t+f_4 \vert +
\frac{f_{3}t+f_4}{{f_{3}(t)x+f_{3}(t)t+f_{4}(t)}}\right]
\]
\begin{equation}
\hskip0.25in -\frac{h_{3}(t)t+h_{4}(t)}{f_{2}\left[f_{2}(t)x+f_{3}(t)t+f_{4}(t)\right]}+
f(t),
\label{non-standard-null-1}
\end{equation}
where
$h_{2}(t):=[\dot{f}_{1}(t)f_{2}(t)-f_{1}(t)\dot{f}_{2}(t)]$, 
$h_{3}(t):=[\dot{f}_{1}(t)f_{3}(t)-f_{1}(t)\dot{f}_{3}(t)]-f_{1}(t)f_{3}(t)$, 
and
$h_{4}(t):=[\dot{f}_{1}(t)f_{4}(t)-f_{1}(t)\dot{f}_{4}(t)]$. 
It is worth mentioning that any other suitably chosen differentiable and integrable 
function $B(x,t)$ of the form in Eq. (\ref{generating-function}) can be used as a 
generating function to construct its corrensponding non-standard null Lagrangian.

Again, the harmonics of any non-standard null-Lagrangian can be obtained using 
the procedure presented in Proposition 2.  For instance, the first harmonics of the 
above non-standard null Lagrangian in Eq. (\ref{non-standard-null-1}) can be 
calculated as 
\[
L^{(1)}_{nsn}(\dot{x}, x,t) = \left[\frac{f_{1}(t)}{f_{2}(t)x+f_{3}(t)t+f_{4}(t)}-
\frac{f_{1}(t)f_{2}(t)}{[f_{2}(t)x+f_{3}(t)t+f_{4}(t)]^2}\right]\dot{x}
\]
\[
\hskip0.25in +\frac{h_{2}(t)}{f_{2}^{2}}\left[ln \vert f_{2}x+f_{3}t+f_4 \vert +
\frac{f_{3}t+f_4}{{f_{3}(t)x+f_{3}(t)t+f_{4}(t)}}\right]
\]
\begin{equation}
\hskip0.25in -\frac{h_{3}(t)t+h_{4}(t)}{f_{2}\left[f_{2}(t)x+f_{3}(t)t+f_{4}(t)\right]}
+\frac{h_{2}(t)x+h_{3}(t)t+h_{4}(t)}{\left[f_{2}(t)x+f_{3}(t)t+f_{4}(t)\right]^2} +f(t).
\label{non-standard-null-harmonics}
\end{equation}
After the exploration of the above strategy for constructing both standard and non-standard 
null Lagrangians, it is natural to seek for similar strategy for constructing non-standard 
Lagrangians. In the following section, it is shown that indeed a simple strategy for 
constructing non-standard Lagrangians from the above null-Lagrangians can be adopted.

\section{From null to non-standard Lagrangians}

The construction of non-standard Lagrangians follows from the availability of the smooth 
and invertible Lagrangians defined as $C^{\infty}$ functions on the domain $\mathcal{U}
\subset{T\mathcal{Q}}$. After a desirable Lagrangian, $L(\dot{x},x,t)$, is identified, a 
non-standard Lagrangian, $\mathcal{L}(\dot{x},x,t)$ may be constructed by function 
composition as $\mathcal{L}(\dot{x},x,t)= (F\circ L)(\dot{x},x,t)=F( L(\dot{x},x,t))$ 
over the same domain $\mathcal{U}\subset{T\mathcal{Q}}$. The following proposition 
encodes this construction of a non-standard Lagrangian and the derivation of its 
corresponding evolution equation.

{\bf Proposition 3:} Let $L(\dot{x},x,t): \mathcal{U} \subset{T\mathcal{Q}}\rightarrow 
\mathbf{R}$ be an invertible and $C^{\infty}$ function. Then a non-standard Lagrangian 
$\mathcal{L}:  \mathcal{U} \subset{T\mathcal{Q}} \rightarrow \mathbf{R}$ can be 
defined by using at least a $C^2$ function $F: \mathbf{R} \rightarrow \mathbf{R}$ 
as $\mathcal{L}(\dot{x},x,t)=(F \circ L)(\dot{x},x,t))=F( L(\dot{x},x,t))$  and the 
corresponding equation of motion is given by
\begin{equation}
p_{L} \left[\frac{d^{2} F}{d L^{2}}\right]\frac{dL}{dt}+ \left(\dot{p}_{L}-f_{L}\right)
\frac{dF}{dL}=0,
\label{proposition-3}
\end{equation}
where $p_{L}:=\partial L/\partial \dot{x}$, $\dot{p}_{L}=dp_{L}/dt$ and $f_{L} :
={\partial{L}}
/{\partial{{x}}}$.

{\bf Proof:} First of all, the composition is well defined since $L(\dot{x},x,t): \mathcal{U} 
\subset{T\mathcal{Q}}\rightarrow \mathbf{R}$ is invertible and $F: \mathbf{R} \rightarrow 
\mathbf{R}$ is at least a $C^{2}$ function  and thereby implying $\mathcal{L}:=(F\circ L):  
\mathcal{U} \subset{T\mathcal{Q}} \rightarrow \mathbf{R}$ is at least a $C^{2}$ function, 
which can therefore be used to define the action functional such as Eq. (\ref{null-L}) for 
extremization. Therefore, the derivation of Eq. (\ref{proposition-3}) follows directly from 
$\hat{EL}[\mathcal{L}=F(L(\dot{x},x,t))]=0$.

It should be noted that no assumption on Lagrangians $L(\dot{x},x,t) \in \mathcal{F}
(\mathcal{U} \subset{T\mathcal{Q}})$  being null is made in the above proposition.  
As a simple check, the equation of motion for $\mathcal{L}=F(L(\dot{x},x,t))=L(\dot{x},
x,t)$ reproduces the expected $\dot{p}_{L}-f_{L}=0$.  Some interesting forms of 
$\mathcal{L}=F(L(\dot{x},x,t))$ are $F(L(\dot{x},x,t))=exp{(L)}$ [11], $F(L(\dot{x},x,t))
=ln{(L)}$ and $F(L(\dot{x},x,t))={L}^{-1}$. Upon inserting these forms in Eq. 
(\ref{proposition-3}), the respective equations of motion are obtained as ${dL}/{dt}+
({\dot{p}_{L}-f_{L}})/{p_L}=0$, ${dL}/{dt}-[({\dot{p}_{L}-f_{L}})/{p_L}]L=0$ 
and ${dL}/{dt}-[({\dot{p}_{L}-f_{L}})/{2p_L}]L=0$.

Next, it is interesting to note that Proposition 3 simplifies considerably for functions of null 
Lagrangians. The use of null Lagrangians in Proposition 3 eliminates the second term in Eq. 
(\ref{proposition-3}), for $\dot{p}_{L_{null}}=f_{L_{null}}$ for all null Lagrangians. In 
particular, inserting $\mathcal{L}=F(L_{null}(\dot{x},x,t))={L}_{null}(\dot{x},x,t)$ in Eq. 
(\ref{proposition-3}) gives the expected $\hat{EL}[\mathcal{L}=L_{null}]=0$ validating 
that no equation of motion is obtained from $\mathcal{L}=L_{null}$. However, most 
importantly, the final equation of motion derived from a function of a null Lagrangian 
turns out to be independent of the form of the function provided that $p_{L} 
\left[\frac{d^{2} F}{dL_{null}^{2}}\right]\ne 0$ for a given function $L_{null}$, 
which is encoded in the following Corollary.

{\bf Corollary 1:} Let $\mathcal{L}:=(F\circ L_{null}):  \mathcal{U} \subset{T\mathcal{Q}} 
\rightarrow \mathbf{R}$ be a set of $C^\infty$ functions composed from  ${L_{null}}:   
\mathcal{U} \subset{T\mathcal{Q}} \rightarrow \mathbf{R}$ and $F: \mathbf{R} 
\rightarrow \mathbf{R}$ \.   Then, for any permissible form of function $\mathcal{L}=
(F\circ L_{null})(\dot{x},x,t)$ satisfying $p_{L_{null}} \left[\frac{d^{2} F}{d L_{null}^{2}}
\right]\ne 0$, the corresponding equation of motion is given by 
\begin{equation}
 \frac{d}{dt}\left[L_{null}\right] =0.
\label{corollary-1}
\end{equation}

The simplicity of the derivation of an equation of motion from Eq. (\ref{corollary-1}) is rather 
remarkable, for it generates an equation of motion describing a dynamical system simply from 
the total time derivative of a permissible null Lagrangian defined over a suitable domain. In 
other words, there exists a dynamical system corresponding to every permissible null 
Lagrangian over a domain of interest. Furthermore, the form of the corresponding equation 
of motion is independent of the functional form defined over different domain for each functional. 
For example, the simplified equation of motion for $F[L_{null}(\dot{x},x,t)]=exp{(L_{null})}$ 
hinges upon the assumption that $p_{L_{null}} \ne 0$ and $L_{null}$ is finite (but can be 
zero at some points) over the chosen domain. On the contrary, both $p_{L_{null}} \ne 0$ 
and ${L_{null}} \ne 0$ must be satisfied over the same domain for $F[L_{null}(\dot{x},x,t)]
=ln{(L_{null})}$. 

At this point, it is desirable to explicitly express the equation of motion for a null Lagrangian in 
the form presented in the previous sections. Therefore, an invertible null Lagrangian, $L(\dot{x},x,t) 
= B(x,t)\dot{x}+C(x,t)x+f(t)$ with its null condition Eq. (\ref{BC-condition}) over a domain 
$\mathcal{U}\subset{T\mathcal{Q}}$, generates the following equation of motion:

\begin{equation}
B(x,t) \ddot{x} +\left[B^{'}(x,t)\dot{x}+2\dot{B}(x,t)\right]\dot{x}+\dot{C}(x,t)x
+\dot{f}(t)=0,
\label{nulleom}
\end{equation}
where primes and dots are defined respectively as $B^{'}:=\partial B(x,t)/\partial x$ and 
$\dot{B} :=\partial B(x,t)/\partial t$. 

Similarly, the equation of motion for $n$-th harmonics of an invertible null Lagrangian,
 $L_{null}(\dot{x},x,t) = B(x,t)\dot{x}+C(x,t)x+f(t)$ over some domain $\mathcal{U} 
\subset{T\mathcal{Q}}$, can be computed from

\begin{equation}
\frac{dL^{(n)}_{null}}{dt} = \frac{dL^{(n-1)}_{null}}{dt} +\frac{d^{2}B_{n-1}(x,t)}
{dt^2} = 0\ ,
\label{eom-harmonics}
\end{equation}
where again $L^{(n-1)}_{null}$ and $B_{n-1}(x,t) $ are defined in Eqs. 
(\ref{null-lagrangians-harmonics}) and (\ref{B_n}) respectively. 

Now, after constructing the non-standard Lagrangians from null Lagrangians and the simplified 
method for obtaining the corresponding equation of motion, it is imperative that the above
 formulation be applied to explore the range of allowed physical systems.  Therefore, a few 
specific examples are presented below to apply these discoveries for a range of allowed 
dynamical systems.

\section{Applications to dynamical systems}

\subsection{Newton's law of inertia}

The general form of the equation of motion (see Eq. \ref{nulleom}) resulting from 
Eq. (\ref{corollary-1}) for the null Lagrangian $L_{null}(\dot{x},x,t) = B(x,t)\dot{x}
+C(x,t)x+f(t)$ may imply that the equation represents a broad variety of dynamical 
systems.  However, it must be kept in mind that there is also the null condition that 
can be written as  
\begin{equation}
\left ( \frac{\partial B(x,t)}{\partial t} \right ) = \frac{\partial [xC(x,t)]}{\partial x}\ ,
\label{eq1}
\end{equation}
and that this condition imposes stringent constraints on admissible functions $B(x,t)$ 
and $C(x,t)$. As a result, only certain dynamical systems may have null Lagrangians 
that allow deriving their equations of motion directly from Eq. (\ref{corollary-1}),
which is a necessary condition for obtaining the equation of motion directly from 
a given null Lagrangian.  The existence of this condition is a new phenomenon in 
the calculus of variations.   

In the simplest case when $B(x,t) = c_1$ = const, $C(x,t) = c_2$ = const, 
and $f(t) = c_3$ = const, Eq. (\ref{nulleom}) reduces to $\ddot{x} = 0$, 
which is the Newton law of inertia.  Then, $L_{null}(\dot{x}) = c_1 \dot x$, 
which is the simplest null Lagrangian.  It is easy to see that substitution of 
this Lagrangian into Eq. (\ref{corollary-1}) gives the required equation of 
motion.  Thus, the law of inertia can be derived by using the null Lagrangian,
which is a new result of this paper.

\subsection{Dissipative systems with constant coefficients}

To indentify dynamical systems, whose equations of motion can be obtained 
from the null Lgrangians by using Eq. (\ref{corollary-1}), we consider the 
following equation 
\begin{equation}
\ddot{x} + \alpha_o \dot{x}^2 + \beta_o \dot{x} + \gamma_o x = 0\ ,
\label{eq2}
\end{equation}
where $\alpha_o$ = const, $\beta_o$ = const and $\gamma_o$ = const.
Comparison of this equation to Eq. (\ref{nulleom}) allows finding the
functions $B(x,t)$ and $C(x,t)$ and obtaining the corresponding null 
Lagrangian.

However, if the coefficients in Eq. (\ref{eq2}) are constant, we obtain  
\begin{equation}
B (x, t) = B_o e^{\alpha_o x + \beta_o t / 2}\ ,
\label{eq3}
\end{equation}
and
\begin{equation}
C (x, t) = 2 B_o \left ( \frac{\gamma_o}{\beta_o} \right ) e^{\alpha_o x 
+ \beta_o t / 2}\ .
\label{eq4}
\end{equation}
Moreover, the null condition gives $(1 + \alpha_o x) \gamma_o = 
\beta_o^2 / 4$.  Since the coefficients are constant, the dependence on
$x$ must be eliminated by either $\alpha_o = 0$ or $\alpha_o \ne 0$ 
but $\beta_o = \gamma_o = 0$.  This shows that there is no null 
Lagrangian for the equation of motion $\ddot x + \gamma_o x = 0$,
which for $\gamma_o$ being a spring constant represents a harmonic 
oscillator.  Our method of deriving equations of motion from null 
Lagrangians seems to be restricted to dissipative systems, with the 
law of inertia being an exception.

We now present the resulting ODEs for the cases when null Lagrangians 
exist and can be used used to derived them.  With $\alpha_o = 0$ and 
$\gamma_o = \beta_o^2 / 4$, Eq. (\ref{eq2}) reduces to 
\begin{equation}
\ddot{x} + \beta_o \dot{x} + \frac{1}{4} \beta_o^2 x = 0\ ,
\label{eq5}
\end{equation}
and its null Lagrangian is 
\begin{equation}
L_{null}(\dot{x},x,t) = \left ( \dot x + \frac{1}{2} x \right ) B_o 
e^{\beta_o t /2}\ .
\label{eq6}
\end{equation}
It is easy to verify that $L_{null}(\dot{x},x,t)$ is the null Lagrangian 
and that this Lagrangian gives the equation of motion (see Eq. 
\ref{eq4}) after it is substituted into Eq. (\ref{corollary-1}). The 
obtained equation of motion describes a damped harmonic oscillator 
in which the damping term and the natural frequency of this system 
are related to each other.

In case $\alpha_o \ne 0$ and $\beta_o = \gamma_o = 0$, the 
resulting equation of motion is 
\begin{equation}
\ddot{x} + \alpha_o \dot{x}^2 = 0\ ,
\label{eq7}
\end{equation}
and the null Lagrangian that gives this equation is 
\begin{equation}
L_{null}(\dot{x},x,t) = B_o \dot {x} e^{\alpha_o x}\ .
\label{eq8}
\end{equation}
In this case the equation of motion describes a dynamical 
system with the quadratic damping and its coefficient 
$\alpha_o$ being arbitrary.

\subsection{Dissipative systems with time-dependent coefficients}

Let us generalize Eq. (\ref{eq2}) and consider the following equation 
\begin{equation}
\ddot{x} + \alpha_1 (t) \dot{x}^2 + \beta_1 (t) \dot{x} + \gamma_1 
(t) x = 0\ ,
\label{eq9}
\end{equation}
where $\alpha_1 (t)$, $\beta_1 (t)$ and $\gamma_1 (t)$ are given 
functions of time.  By comparing this equation to Eq. (\ref{nulleom}),
the functions $B(x,t)$ and $C(x,t)$ are obtained
\begin{equation}
B (x, t) = B_o e^{\alpha_1 (t) x + I_{\beta} (t)}\ ,
\label{eq10}
\end{equation}
where
\begin{equation}
I_{\beta} (t) = \frac{1}{2} \int^{t} \beta_1 ({\tilde t}) d \tilde t\ ,
\label{eq11}
\end{equation}
and
\begin{equation}
C (x, t) = B_o \int^{t} \gamma_1 ({\tilde t}) e^{\alpha_1 ({\tilde t}) 
x + I_{\beta} ({\tilde t})} d \tilde t\ .
\label{eq12}
\end{equation}

Then, the null condition is
\[
\left [ \dot \alpha_1 (t) x + \frac{1}{2} \beta_1 (t) \right ] e^{\alpha_1 (t) x 
+ I_{\beta} (t)} = \int^{t} \gamma_1 ({\tilde t}) e^{\alpha_1 ({\tilde t}) 
x + I_{\beta} ({\tilde t})} d \tilde t
\]
\begin{equation}
\hskip2.05in\ + x \int^{t} \alpha_1 ({\tilde t}) \gamma_1 ({\tilde t}) e^{\alpha_1 
({\tilde t}) x + I_{\beta} ({\tilde t})} d \tilde t\ .
\label{eq13}
\end{equation}
Since the coefficients are only functions of $t$, the dependence on $x$ must 
be eliminated.  This can be achieved by taking either $\alpha_1 (t) = 0$ or
$\alpha_1 (t) \neq 0$ but $\beta_1 (t) = \gamma_1 (t) = 0$.  Both cases 
are similar to those considered in Section 5.1 but the resulting equations 
of motion and the null Lagrangians are more general, thus, we present 
them now. 

Having obtained $B (x, t)$ and $C (x, t)$, we take $\alpha_1 (t) = 0$  
and find the following null Lagrangian
\begin{equation}
L_{null}(\dot{x},x,t) = B_o \left [ \dot x e^{I_{\beta} (t)} + x \int^{t} 
\gamma_1 ({\tilde t}) e^{I_{\beta} ({\tilde t})} d \tilde t \right ]\ .
\label{eq14}
\end{equation}
Substituting this Lagrangian into Eq. (\ref{corollary-1}), we obtain 
\begin{equation}
\ddot{x} + \beta_1 (t) \dot{x} + \gamma_1 (t) x = 0\ ,
\label{eq15}
\end{equation}
which is the correct equation of motion.  However, the functions
$\beta_1 (t)$ and $\gamma_1 (t)$ are related to each other 
through the null condition (see Eq. \ref{eq13}) given by
\begin{equation} 
\frac{1}{2} \beta_1 (t) e^{I_{\beta} (t)} = \int^{t} 
\gamma_1 ({\tilde t}) e^{I_{\beta} ({\tilde t})} d \tilde t\ .
\label{eq16}
\end{equation}
This equation generalizes that given by Eq. (\ref{eq5}) and 
describes a damped harmonic oscillator, whose coefficients 
are functions of $t$ and they are dependent.

Now, let us consider the second case of $\beta_1 (t) = \gamma_1
(t) = 0$, which reduces the null condition given by Eq. (\ref{eq13}) 
to $\dot \alpha_1 (t) = 0$ or $\alpha_1 (t) = \alpha_o$ = const.
This is the same case as considered in Section 5.1 and the resulting 
equation of motion and its null Lagrangian are given by Eqs (\ref{eq7})
and (\ref{eq8}), respectively.  The obtained result shows that the 
null Lagrangian exists only for the equation of motion with the 
quadratic damping when the damping coefficient is constant.

\subsection{Dissipative systems with displacement-dependent coefficients}

We may also generalize Eq. (\ref{eq2}) and consider the following equation 
\begin{equation}
\ddot{x} + \alpha_2 (x) \dot{x}^2 + \beta_2 (x) \dot{x} + \gamma_2
(x) x = 0\ ,
\label{eq17}
\end{equation}
where $\alpha_2 (x)$, $\beta_2 (x)$ and $\gamma_2 (x)$ are given 
functions of displacement.  To find $B(x,t)$ and $C(x,t)$, we compare 
the above equation to Eq. (\ref{nulleom}), and obtain
\begin{equation}
B (x, t) = B_o e^{I_{\alpha} (x) + \beta_2 (x) t / 2}\ ,
\label{eq18}
\end{equation}
where
\begin{equation}
I_{\alpha} (x) = \int^{t} \alpha_2 ({\tilde x}) d \tilde x\ ,
\label{eq19}
\end{equation}
and
\begin{equation}
C (x, t) = 2 B_o \frac{\gamma_2 (x)}{\beta_2 (x)} e^{I_{\alpha} (x) 
+ \beta_2 (x) t / 2}\ .
\label{eq20}
\end{equation}

Then, the null condition can be written as 
\begin{equation}
\frac{1}{4} \frac{\beta_2^2 (x)}{\gamma_2 (x)} = 1 + \left [ 
\frac{\gamma^{\prime}_2 (x)}{\gamma_2 (x)} - \frac{\beta^{\prime}_2 
(x)}{\beta_2 (x)} + \alpha_2 (x) + \frac{1}{2} \beta^{\prime}_2 (x) t
\right ]\ .
\label{eq21}
\end{equation}
Since the coefficients depend only on $x$, the dependence on $t$ must 
be eliminated by taking $\beta^{\prime}_2 (x) = 0$, which gives 
$\beta_2 (x) = \beta_o$ = const.  As a result, the null 
condition reduces to 
\begin{equation}
\frac{1}{4} \beta_o^2 = x \gamma^{\prime}_2 (x) + \gamma_2 (x) 
\left [ 1 + \alpha_2 (x) x \right ]\ .
\label{eq22}
\end{equation}

The most general case when the above null condition is satisfied requires 
$\alpha_2 (x)$, $\beta_2 (x) = \beta_o$ and $\gamma_2 (x)$. Then, the 
resulting null Lagrangian is 
\begin{equation}
L_{null}(\dot{x},x,t) = B_o \left [ \dot x + 2 x \frac{\gamma_2 (x)}{\beta_o}
\right ] e^{I_{\alpha} (x) + \beta_o t /2}\ ,
\label{eq23}
\end{equation}
which gives the following equation of motion
\begin{equation}
\ddot{x} + \alpha_2 (x) \dot{x}^2 + \beta_o \dot{x} + \gamma_2 (x) x = 0\ .
\label{eq24}
\end{equation}
In this equation of motion, the functions $\alpha_2 (x)$ and $\gamma_2 (x)$
are related to each other by the null condition (see Eq. \ref{eq22}), which 
also depends on the damping coefficient $\beta_o$.

There are also three special cases: $\alpha_2 (x) = \alpha_o / x$, $\alpha_2 (x) = 0$
and $\beta_o = 0$.  In the first case, we find
\begin{equation}
\gamma_2 (x) = \frac{1}{4} \frac{\beta_o^2}{1 + \alpha_o} + \tilde c_1 x^{-(1+
\alpha_o)}\ ,
\label{eq25}
\end{equation}
where $\tilde c_1$ is an integration constant.  The null Lagrangian is the same as that 
given by Eq. (\ref{eq23}) but with the replacement $\alpha_2 (x) = \alpha_o / x$.
The same replacement in Eq. (\ref{eq24}) gives the required equation of motion.
In the second case, $\alpha_2 (x) = 0$, and the solution to the null condition 
can be found and written as 
\begin{equation}
\gamma_2 (x) = \frac{\tilde c_2}{x} + \frac{1}{4} \beta_o^2\ .
\label{eq25}
\end{equation}
where $\tilde c_2$ is an integration constant.  It is seen that $\gamma_2 (x)$ must 
have a special dependence on $x$ and on the coefiicient $\beta_o$.  The null Lagrangian 
and the resulting equation of motion are given by Eqs (\ref{eq23}) and (\ref{eq24}),
respectively, by taking $\alpha_2 (x) = 0$. 

The third case requires that $\beta_o = 0$, which gives 
\begin{equation}
B (x,t) = B_o e^{I_{\alpha} (x)}\ ,
\label{eq26}
\end{equation}
and 
\begin{equation}
C (x,t) = B_o t \gamma_2 (x) e^{I_{\alpha} (x)}\ .
\label{eq26}
\end{equation}
Then, the null Lagrangian becomes 
\begin{equation}
L_{null}(\dot{x},x,t) = B_o \left [ \dot x + x t \gamma_2 (x) \right ]
e^{I_{\alpha} (x)}\ ,
\label{eq27}
\end{equation}
and the equation of motion is 
\begin{equation}
\ddot{x} + \alpha_2 (x) \dot{x}^2 + \gamma_2 (x) x = 0\ .
\label{eq28}
\end{equation}
Again, the functions $\alpha_2 (x)$ and $\gamma_2 (x)$ are 
not arbitrary but instead they must satisfy the following condition
\begin{equation}
\gamma_2 (x) = \frac{\tilde c_3}{x} e^{I_{\alpha} (x)} \ ,
\label{eq29}
\end{equation}
where $\tilde c_3$ is an integration constant.

\subsection{Comparison to other Lagrangians} 

The presented results show that null Lagrangians can be obtained 
for the law of inertia and some dynamical systems.  An interesting 
result is that, except the first Newton law, these systems must be 
dissipative, which suggests a novel role of null Lagrangians in 
classical dynamics of dissipative systems.  For some systems 
considered in this paper, standard and nonstandard Lagrangians 
were already obtained, so now we compare those Lagrangians 
to the null Lagrangians derived in this paper.

For the law of inertia, $\ddot x = 0$, the standard [1-4] and 
non-standard [28] Lagrangians are given by 
\begin{equation}
L_{sd} (\dot{x}) = \frac{1}{2} \dot x^2\ ,
\label{eq30}
\end{equation}
and 
\begin{equation}
L_{nsd}[\dot{x},x,t]= \frac{1}{C_1(a_o t+v_o)^2[(a_o t+v_o)
\dot{x} - a_o x +C_2]}\ ,
\label{eq31}
\end{equation}
where $C_1$ and $C_2$ being constants of integration, and $v_o$ and
$a_o$ being specified by the initial conditions for solving the auxiliary 
differential equation [28].  Both Lagrangians, when substituted into the 
E-L equation, give the law of inertia.

A new result of this paper is that the null Lagrangian
\begin{equation}
L_{null}(\dot{x}) = c_1 \dot x\ ,
\label{eq32}
\end{equation}
where $c_1$ is a constant, gives also the law of inertia after it is substituted 
into Eq. (\ref{corollary-1}).  Thus, the law of inertia can be derived by using 
either the standard or non-standard or null Lagrangian.  Among these three 
Lagrangians, the simplest one is the null Lagrangian and the most complicated 
one is the non-standard Lagrangian, yet all of them give the same law of inertia. 
The advantage of the Lagrangian formulation presented in this paper is that 
the law of inertia is derived from the null Lagrangian by simply taking its time 
derivative.   

The equation of motion with the quadratic damping term, $\ddot x + \alpha_o
\dot x^2 = 0$, was also previously studied [7,8], and the following standard 
and non-standard Lagrangians were obtained
\begin{equation}
L_{sd} (\dot{x}) = \frac{1}{2} \dot x^2 e^{2 \alpha_o x}\ ,
\label{eq33}
\end{equation}
and 
\begin{equation}
L_{nsd}[\dot{x},x,t]= \frac{1}{\dot{x} e^{\alpha_o x} + 1}\ .
\label{eq34}
\end{equation}
Both Lagrangians, when substituted into the E-L equation, give the equation 
of motion.   The same equation of motion is obtained when the null Lagrangian
\begin{equation}
L_{null}(\dot{x}) = c_2 \dot x e^{\alpha_o x}\ ,
\label{eq35}
\end{equation}
where $c_2$ is a constant, is substituted into Eq. (\ref{corollary-1}).  Comparison 
of the above Lagrangians shows that each one of them contains $\dot x$ and 
$e^{\alpha_o x}$, and that their forms are similar.  

Finally, we consider the equation of motion for a damped oscillator, $\ddot x + 
\beta_o \dot x + \frac{1}{4} \beta^2_o x = 0$, whose standard and non-standard 
Lagrangians [7,8] are 
\begin{equation}
L_{sd} (\dot{x},t) = \frac{1}{2} \left ( \dot x^2 - \frac{1}{4} \beta_o^2 x^2
\right ) e^{\beta_o t}\ ,
\label{eq36}
\end{equation}
and 
\begin{equation}
L_{nsd}[\dot{x},x,t]= \frac{e^{-\beta_o x / 2}}{\dot{x}  + \frac{1}{2} 
\beta_o x}\ .
\label{eq37}
\end{equation}
The null Lagrangian for the equation of motion is 
\begin{equation}
L_{null}(\dot{x},x,t) = c_3 \left (\dot x + \frac{1}{2} \beta_o x \right ) 
e^{\beta_o t / 2}\ .
\label{eq38}
\end{equation}
where $c_2$ is a constant.  Comparison of $L_{nsd} (\dot{x},x,t)$ and 
$L_{null}(\dot{x},x,t)$ shows that the non-standard Lagrangian is exactly 
the inverse of the null Lagrangian, if $c_3 =1$.

\section{Conclusions} 

General null Lagrangians and their higher harmonics are constructed 
for dynamical systems with one degree of freedom.  Since the null
Lagrangians satisfy identically the E-L equation, they cannot be 
directly used to obtain equations of motion.  However, as the 
results of this paper show, the derived null Lagrangians can be 
used to construct non-standard Lagrangians that do give equations 
of motion.  The main result of this paper is the following condition
\begin{equation}
 \frac{d}{dt}\left[L_{null}\right] =0,
\label{main}
\end{equation}
which gives an equation of motion for any null Lagrangian.  Thus,
the condition plays the same role for null Lagrangians as the E-L 
equation plays for standard or non-standard Lagrangians.   

The presented results show how to determine a null Lagrangian 
for a given dynamical system and then how to use this Lagrangian
to obtain the resulting equation of motion.  Specific applications of
the method to general dynamical systems show that the method 
preferentially works for dissipative systems but it also can be used 
to some conservative systems such as the law of inertia.  The results 
of this paper demonstrate a novel role of null Lagrangians in classical 
dynamics.
%



\end{document}